\pdfoutput=1

\newif\ifwordcount
\wordcountfalse

\documentclass[aps,prl,twocolumn,showpacs,superscriptaddress,groupedaddress,titlepage,nofootinbib]{revtex4}
\usepackage{graphicx}
\usepackage{dcolumn}
\usepackage{bm}           
\usepackage{amssymb}
\usepackage{amsmath}
\usepackage{color}
\usepackage{natbib}
\usepackage{enumerate}
\usepackage[svgnames]{xcolor}
\usepackage{aas_macros}

\usepackage[breaklinks, colorlinks, citecolor=blue, colorlinks=true, linkcolor=blue, urlcolor=blue]{hyperref}
\usepackage{ifthen}

\def\eprinttmp@#1arXiv:#2 [#3]#4@{
\ifthenelse{\equal{#3}{x}}{\href{http://arxiv.org/abs/#1}{#1}}{\href{http://arxiv.org/abs/#2}{arXiv:#2} [#3]}}
\providecommand{\eprint}[1]{\eprinttmp@#1arXiv: [x]@}
\newcommand{\adsurl}[1]{\href{#1}{ADS}}

\def\be{\begin{equation}}
\def\ee{\end{equation}}
\def\ba{\begin{eqnarray}}
\def\ea{\end{eqnarray}}

\def\muKarcmin{\mu \mathrm{K}\mbox{-}{\rm arcmin}}
\def\arcmin{{'}}

\def\micron{\mu \mathrm{m}}

\begin{document}

\title{Detection of $B$-mode Polarization in the Cosmic Microwave Background \\ with Data from the South Pole Telescope}

\def\affMcGill{\affiliation{Department of Physics, McGill University, Montreal, Quebec H3A 2T8, Canada}}
\def\affKICP{\affiliation{Kavli Institute for Cosmological Physics, University of Chicago, Chicago, Illinois 60637, USA}}
\def\affChicagoP{\affiliation{Department of Physics, University of Chicago, Chicago, Illinois 60637, USA}}
\def\affChicagoA{\affiliation{Department of Astronomy and Astrophysics, University of Chicago, Chicago, Illinois 60637, USA}}
\def\affChicago{\affiliation{University of Chicago, Chicago, Illinois 60637, USA}}
\def\affColoradoA{\affiliation{Department of Astrophysical and Planetary Sciences and Department of Physics, University of Colorado, Boulder, Colorado 80309, USA}}
\def\affNIST{\affiliation{National Institute of Standards and Technology, Boulder, Colorado 80305, USA}}
\def\affEFI{\affiliation{Enrico Fermi Institute, University of Chicago, Chicago, Illinois 60637, USA}}
\def\affArgonneH{\affiliation{High Energy Physics Division, Argonne National Laboratory, Argonne, Illinois 60439, USA}}
\def\affCaltech{\affiliation{California Institute of Technology, Pasadena, California 91125, USA}}
\def\affJPL{\affiliation{Jet Propulsion Laboratory, Pasadena, California 91109, USA}}
\def\affKwaZulu{\affiliation{School of Mathematics, Statistics \& Computer Science, University of KwaZulu-Natal, Durban 4000, South Africa}}
\def\affColoradoP{\affiliation{Department of Physics, University of Colorado, Boulder, Colorado 80309, USA}}
\def\affMinnesota{\affiliation{Department of Physics, University of Minnesota, Minneapolis, Minnesota 55455, USA}}
\def\affBerkeley{\affiliation{Department of Physics, University of California, Berkeley, California 94720, USA}}
\def\affLBNL{\affiliation{Physics Division, Lawrence Berkeley National Laboratory, Berkeley, California 94720, USA}}
\def\affBCCP{\affiliation{Berkeley Center for Cosmological Physics, Department of Physics, University of California, and Lawrence Berkeley National Laboratory, Berkeley, California 94720, USA}}
\def\affArgonneM{\affiliation{Materials Science Division, Argonne National Laboratory, Argonne, Illinois 60439, USA}}
\def\affArgonne{\affiliation{Argonne National Laboratory, Argonne, Illinois 60439, USA}}
\def\affDavis{\affiliation{Department of Physics, University of California, Davis, California 95616, USA}}
\def\affUBC{\affiliation{Department of Physics and Astronomy, University of British Columbia, Vancouver, British Columbia V6T 1Z1, Canada}}
\def\affMichigan{\affiliation{Department of Physics, University of Michigan, Ann  Arbor, Michigan 48109, USA}}
\def\affCase{\affiliation{Physics Department, Case Western Reserve University, Cleveland, Ohio 44106, USA}}
\def\affTorontoD{\affiliation{Dunlap Institute for Astronomy and Astrophysics, University of Toronto, 50 St. George Street, Toronto, Ontario M5S 3H4, Canada}}
\def\affTorontoA{\affiliation{Department of Astronomy and Astrophysics, University of Toronto, 50 St. George Street, Toronto, Ontario M5S 3H4, Canada}}
\def\affCfa{\affiliation{Harvard-Smithsonian Center for Astrophysics, Cambridge, Massachusetts 02138, USA}}
\def\affIPAC{\affiliation{Infrared Processing and Analysis Center, California Institute of Technology, JPL, Pasadena, California 91125, USA}}
\def\affArt{\affiliation{Liberal Arts Department, School of the Art Institute of Chicago, Chicago, Illinois 60603, USA}}
\def\affCardiff{\affiliation{School of Physics and Astronomy, Cardiff University, Cardiff CF24 3YB, United Kingdom}}
\def\affColoradoG{\affiliation{CASA, Department of Astrophysical and Planetary Sciences, University of Colorado, 389 UCB, Boulder, Colorado 80309, USA}}

\author{
\renewcommand*\thefootnote{\textcolor{blue}{*}}
D.~Hanson{\footnote{\href{mailto:dhanson@physics.mcgill.ca}{\color{blue}{Electronic address: dhanson@physics.mcgill.ca}}}}}
\affMcGill

\author{S.~Hoover}
\affKICP
\affEFI

\author{A.~Crites}
\affKICP
\affChicagoA

\author{P.~A.~R.~Ade}
\affCardiff

\author{K.~A.~Aird}
\affChicago

\author{J.~E.~Austermann}
\affColoradoG

\author{J.~A.~Beall}
\affNIST

\author{A.~N.~Bender}
\affMcGill

\author{B.~A.~Benson}
\affKICP
\affEFI

\author{L.~E.~Bleem}
\affKICP
\affChicagoP

\author{J.~J.~Bock}
\affCaltech
\affJPL

\author{J.~E.~Carlstrom}
\affKICP
\affEFI
\affChicagoA
\affChicagoP
\affArgonneH

\author{C.~L.~Chang}
\affArgonneH
\affKICP
\affEFI

\author{H.~C.~Chiang}
\affKICP
\affKwaZulu

\author{H-M.~Cho}
\affNIST
\affColoradoG

\author{A.~Conley}
\affColoradoG

\author{T.~M.~Crawford}
\affKICP
\affChicagoA

\author{T.~de~Haan}
\affMcGill

\author{M.~A.~Dobbs}
\affMcGill

\author{W.~Everett}
\affColoradoG

\author{J.~Gallicchio}
\affKICP

\author{J.~Gao}
\affNIST

\author{E.~M.~George}
\affBerkeley

\author{N.~W.~Halverson}
\affColoradoG
\affColoradoP

\author{N.~Harrington}
\affBerkeley

\author{J.~W.~Henning}
\affColoradoG

\author{G.~C.~Hilton}
\affNIST

\author{G.~P.~Holder}
\affMcGill

\author{W.~L.~Holzapfel}
\affBerkeley

\author{J.~D.~Hrubes}
\affChicago

\author{N.~Huang}
\affBerkeley

\author{J.~Hubmayr}
\affNIST

\author{K.~D.~Irwin}
\affNIST

\author{R.~Keisler}
\affKICP
\affChicagoP

\author{L.~Knox}
\affDavis

\author{A.~T.~Lee}
\affBerkeley
\affLBNL

\author{E.~Leitch}
\affKICP
\affChicagoA

\author{D.~Li}
\affNIST

\author{C.~Liang}
\affKICP
\affChicagoA

\author{D.~Luong-Van}
\affKICP

\author{G.~Marsden}
\affUBC

\author{J.~J.~McMahon}
\affMichigan

\author{J.~Mehl}
\affKICP
\affArgonneH

\author{S.~S.~Meyer}
\affKICP
\affChicagoP
\affEFI
\affChicagoA

\author{L.~Mocanu}
\affKICP
\affChicagoA

\author{T.~E.~Montroy}
\affCase

\author{T.~Natoli}
\affKICP
\affChicagoP

\author{J.~P.~Nibarger}
\affNIST

\author{V.~Novosad}
\affArgonneM

\author{S.~Padin}
\affCaltech

\author{C.~Pryke}
\affMinnesota

\author{C.~L.~Reichardt}
\affBerkeley

\author{J.~E.~Ruhl}
\affCase

\author{B.~R.~Saliwanchik}
\affCase

\author{J.~T.~Sayre}
\affCase

\author{K.~K.~Schaffer}
\affKICP
\affArt

\author{B.~Schulz}
\affCaltech
\affIPAC

\author{G.~Smecher}
\affMcGill

\author{A.~A.~Stark}
\affCfa

\author{K.~T.~Story}
\affKICP
\affChicagoP

\author{C.~Tucker}
\affCardiff

\author{K.~Vanderlinde}
\affMcGill
\affTorontoD
\affTorontoA

\author{J.~D.~Vieira}
\affCaltech

\author{M.~P.~Viero}
\affCaltech

\author{G.~Wang}
\affArgonneH

\author{V.~Yefremenko}
\affArgonneH
\affArgonneM

\author{O.~Zahn}
\affBCCP

\author{M.~Zemcov}
\affCaltech
\affJPL

\date{30 September 2013}

\begin{abstract}
Gravitational lensing of the cosmic microwave background generates a curl pattern in the observed polarization.
This ``$B$-mode'' signal provides a measure of the projected mass distribution over the entire observable Universe
and also acts as a contaminant for the measurement of primordial gravity-wave signals.
In this Letter we present the first detection of gravitational lensing $B$ modes, 
using first-season data from the polarization-sensitive receiver on the South Pole Telescope (SPTpol).
We construct a template for the lensing $B$-mode signal by combining
$E$-mode polarization measured by SPTpol with estimates of the lensing potential from a 
\textit{Herschel}-SPIRE map of the cosmic infrared background.
We compare this template to the $B$ modes measured directly by SPTpol,
finding a non-zero correlation at $7.7\sigma$ significance.
The correlation has an amplitude and scale-dependence consistent with theoretical 
expectations, is robust with respect to analysis choices, and constitutes the 
first measurement of a powerful cosmological observable.
\end{abstract}

\ifwordcount
\else
\maketitle
\fi

\textbf{Introduction:}
Maps of the cosmic microwave background (CMB) 
\cite{Hu:2001bc}
polarization anisotropies
are naturally decomposed into 
curl-free $E$ modes and 
gradient-free $B$ modes
\cite{Kamionkowski:1996ks,Zaldarriaga:1996xe}.
$B$ modes are not generated at linear order in perturbation theory by the scalar perturbations which are the dominant
source of CMB temperature and $E$-mode anisotropies.
Because of this, $B$ modes are of great interest as a clean
probe of two more subtle signals:
(1) primordial tensor perturbations in the early Universe \cite{Kamionkowski:1996zd,Seljak:1996gy},
the measurement of which would provide a unique probe of the energy scale of inflation; and
(2) gravitational lensing, 
which generates a distinctive non-Gaussian $B$-mode signal \cite{Zaldarriaga:1998ar} that 
can be used to measure the projected mass distribution and 
constrain cosmological parameters such as the sum of neutrino masses (for a review, see \cite{Smith:2008an}).

Previous experiments have placed upper limits on the $B$-mode polarization anisotropy 
\cite{Chiang:2009xsa,Brown:2009uy,Bennett:2012zja,Araujo:2012yh}.
In this Letter we present the first detection of $B$ modes sourced by gravitational lensing, 
using first-season data from SPTpol, the polarization-sensitive receiver on the South Pole Telescope.

Gravitational lensing remaps the observed position of CMB anisotropies as 
$\hat{n} \rightarrow \hat{n} + \nabla \phi(\hat{n})$,
where $\phi$ is the CMB lensing potential \cite{Lewis:2006fu}.
This remapping mixes some of the (relatively) large $E$-mode signal into $B$.
The induced $B$ mode at Fourier wavevector $\vec{l}_B$ 
is given to first order in $\phi$ as \cite{Hu:2001kj}
\be
B^{\rm lens}(\vec{l}_B) = \int d^2 \vec{l_E} \int d^2 \vec{l_\phi} W^{\phi}(\vec{l}_E, \vec{l_B}, \vec{l}_{\phi}) E(\vec{l}_E) \phi(\vec{l}_\phi),
\label{eqn:blen}
\ee
where the weight function $W^{\phi}$ specifies the mixing.
In this Letter, we use measurements of $E$ and $\phi$ to synthesize an estimate for the lensing contribution,
which we cross-correlate with measured $B$ modes.
Using maps of the cosmic infrared background measured by the SPIRE instrument onboard the \textit{Herschel} space observatory 
to estimate $\phi$,
and measurements of the $E$- and $B$-mode polarization from SPTpol,
we detect the lensing signal at $7.7\sigma$ significance.

\textbf{CMB Data:}
The South Pole Telescope (SPT) \cite{Carlstrom:2009um} is a 10-meter telescope located at the geographic South Pole.
Here we use data from SPTpol, a polarization-sensitive receiver installed on the telescope in January 2012.
SPTpol
consists of two arrays of polarization-sensitive bolometers (PSBs):
1176\,PSBs that observe at 150\,GHz, and 360\,PSBs that observe at 95\,GHz.
The instrument and its performance are described in Refs.~\cite{george12,austermann12,sayre12,henning12}.
The observation strategy, calibration, and data reduction for SPTpol data are similar to those used for 
the SPT-SZ survey, described in Ref.~\cite{story12b}.
Here we briefly summarize the important points.

We calibrate the PSB polarization sensitivities with observations of a 
ground-based thermal source behind a polarizing grid. 
This allows us to measure the polarization angle of individual PSBs
with $<\!2^{\rm o}$ statistical uncertainty and the average angle
of all PSBs with $<\!0.1^{\rm o}$ statistical uncertainty.
We estimate systematic uncertainty on the average angle to be 
$<\!1^{\rm o}$ ($1.5^{\rm o}$) at 150\,GHz (95\,GHz).

Between March and November 2012, we used SPTpol to observe a $100\,{\rm deg}^2$ region of low-foreground sky,
between 23h and 24h in right ascension and $-50$ and $-60$~deg in declination.
We process the SPTpol data by ``observations'', which are half-hour periods in which the 
telescope scans half of the field.
Each observation is recorded as time-ordered data (TOD) from each PSB, in azimuthal scans separated by steps in elevation.
For each scan, we apply a low-pass anti-aliasing filter as well as a high-pass 
4th-order polynomial subtraction to remove large-scale atmospheric fluctuations. 
This suppresses modes along the scan direction,
which we account for with a two-dimensional transfer function measured from simulations 
of the filtering process. 

In each observation, we drop PSBs with cuts  based on noise level during the observation, 
response to elevation-dependent atmospheric power, and response to an internal thermal calibration source. 
Typical observations include $\sim\!800$~PSBs ($\sim\!230$~PSBs) at 150\,GHz (95\,GHz). 
We cut scans for PSBs with glitches (caused, for example, by cosmic ray hits). 
In typical 150\,GHz (95\,GHz) observations, we lose $\sim\!1\%$ ($\sim\!4\%$) of the data due to glitch removal.

Data from each PSB are accumulated into 
maps of the I, Q, and U Stokes parameters using 
measured polarization angles and polarization efficiencies.
We weight the TOD for each PSB in a scan by the inverse of the variance along the scan direction between 1\,Hz and 3\,Hz 
\mbox{($1300\lesssim l_x \lesssim 3900$} for the telescope scan speed of \mbox{$0.28$ deg/s}).
We make maps using the oblique Lambert azimuthal equal-area projection 
\cite{snyder87} with square $2\arcmin\!\times\!2\arcmin$ pixels.
This projection preserves area on the sky but introduces small distortions in angle;
we account for these distortions by rotating the Q and U components to maintain a 
consistent angular coordinate system across the map.
For each observation we form a noise map from the difference of left- and right-going scans, 
cutting observations which are outliers in metrics such as overall variance.
This cut removes $\sim\!8\%$ ($\sim\!9\%$) of the 150\,GHz (95\,GHz) data. 
Finally, we add the individual observations together to produce full-season maps, with polarization noise levels
of approximately $10\,\muKarcmin$ at 150\,GHz and $25\,\muKarcmin$ at 95\,GHz.

\begin{figure*}[!tb]
\centerline{\includegraphics[width=\textwidth]{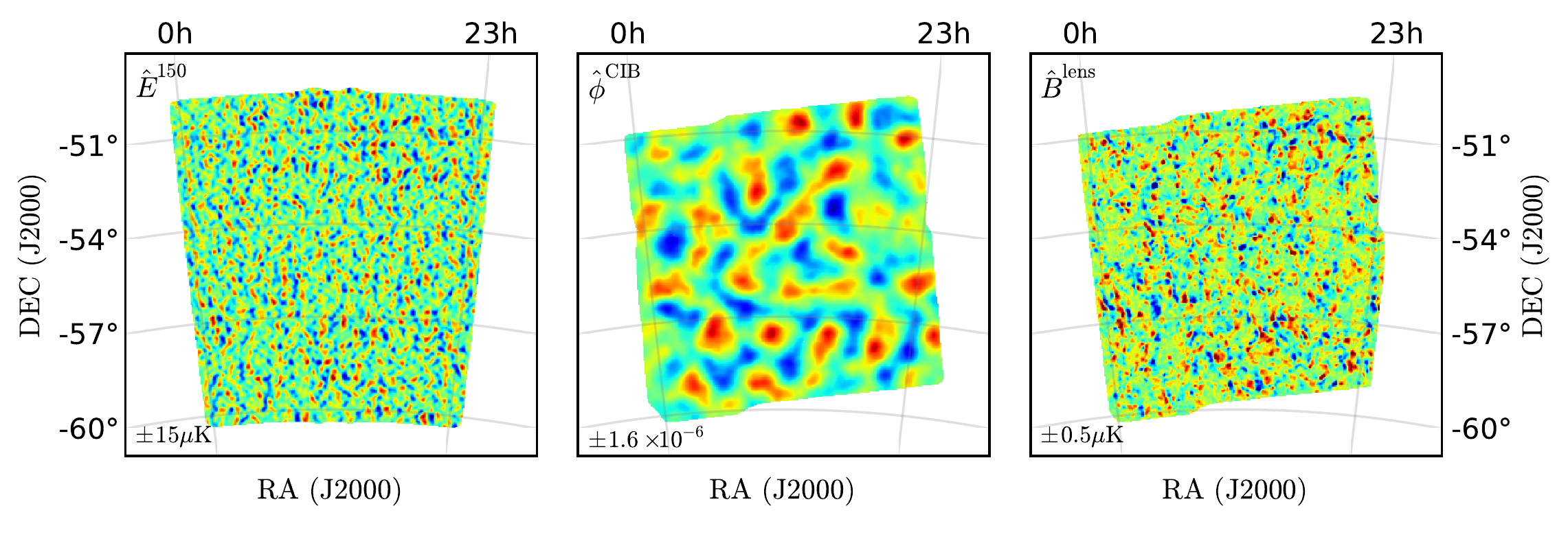}}
\vspace{-0.2in}
\caption{
Left: Wiener-filtered $E$-mode polarization measured by SPTpol at $150$\,GHz.
Center: Wiener-filtered CMB lensing potential inferred from CIB fluctuations measured by \textit{Herschel} at $500\,\micron$.
Right: gravitational lensing $B$-mode estimate synthesized using Eq.~\eqref{eqn:blen}.
The lower left corner of each panel indicates the blue(-)/red(+) color scale.
}
\label{fig:map_ebp_tri}
\end{figure*}

Inaccuracy in PSB gain measurements can cause direct leakage of the CMB temperature into polarization,
which we fit for using the cross-spectra of I with \mbox{Q and U}.
We find $<\!2\%$ leakage at both $150$\,GHz and $95$\,GHz, which we correct for by subtracting 
appropriate fractions of I from Q and U. 
We show below that this correction is unimportant for our final results.

We calibrate the overall amplitude of the SPTpol maps to better than $1 \%$ in temperature by cross correlating 
with SPT-SZ temperature maps over the same region of sky. 
The SPT-SZ maps are calibrated by comparing to the \textit{Planck} surveyor 143 GHz 
maps \cite{Ade:2013ktc} over the full 2500\,$\mathrm{deg}^2$ SPT-SZ survey region. 

\textbf{CIB Data:}
We use maps of the cosmic infrared background (CIB)
\cite{Holder:2013hqu} obtained from the SPIRE instrument \cite{Griffin:2010hp}
onboard the \textit{Herschel} space observatory \cite{Pilbratt:2010mv} as a tracer of the
CMB lensing potential $\phi$.
The CIB has been established as a well-matched tracer of the lensing potential \cite{Song:2002sg,Holder:2013hqu,Ade:2013aro} and 
currently provides a higher signal-to-noise estimate of $\phi$ than is available with CMB lens reconstruction.
Its use in cross-correlation with the SPTpol data also makes our measurement less sensitive to instrumental systematic effects \cite{Benabed:2000jt}.
We focus on the \textit{Herschel} $500\,\micron$ map, which has the best overlap with the CMB lensing kernel \cite{Holder:2013hqu}.

\textbf{Post-Map Analysis:}
We obtain Fourier-domain CMB temperature and polarization modes using a
Wiener filter (for example, Ref.~\cite{Elsner:2012fe} and references therein), derived
by maximizing the likelihood of the observed I, Q, and U maps as a function of the fields 
$T(\vec{l})$, $E(\vec{l})$, and $B(\vec{l})$.
The filter simultaneously deconvolves the two-dimensional transfer function due to the beam, TOD filtering,
and map pixelization while down-weighting modes that are ``noisy'' due to either atmospheric fluctuations, 
extragalactic foreground power, or instrumental noise.
We place a prior on the CMB auto-spectra, using the best-fit cosmological model given by Ref.~\cite{Ade:2013zuv}.
We use a simple model for the extragalactic foreground power in temperature \cite{story12b}.
We use jackknife difference maps to determine a combined atmosphere+instrument noise model, following Ref.~\cite{vanEngelen:2012va}.
We set the noise level to infinity for any pixels within
$5\arcmin$ of sources detected at $>5\sigma$ in \cite{mocanu13}.
We extend this mask to $10\arcmin$ for all sources with flux greater than $50\,{\rm mJy}$, as
well as galaxy clusters detected using the
Sunyaev-Zel'dovich effect
 in Ref.~\cite{vanderlinde10}.
These cuts remove approximately $5\,{\rm deg}^2$ of the total $100\,{\rm deg}^2$ survey area.
We remove spatial modes close to the scan direction with an $\ell_x < 400$ cut, as well as all modes with $l > 3000$.
For these cuts, our estimated beam and filter map transfer functions are within $20\%$ of unity for every unmasked mode 
(and accounted for in our analysis in any case).

The Wiener filter naturally separates $E$ and $B$ contributions, although in principle this separation depends on the priors placed on their power spectra.
To check that we have successfully separated $E$ and $B$, we also 
form a simpler estimate using the $\chi_B$ formalism advocated in \cite{Smith:2006vq}.
This uses numerical derivatives to estimate a field $\chi_B(\vec{x})$ which is proportional to $B$ in harmonic space. 
This approach cleanly separates $E$ and $B$, although it can be somewhat noisier due to mode-mixing induced by point source masking.
We therefore do not mask point sources when applying the $\chi_B$ estimator.

We obtain Wiener-filtered estimates $\hat{\phi}^{\rm CIB}$ of the lensing potential from the \textit{Herschel} $500\,\micron$ maps by applying an apodized mask, Fourier transforming, and then multiplying by 
$C_l^{{\rm CIB}\mbox{-}\phi} ( C_l^{{\rm CIB}\mbox{-}{\rm CIB}} C_l^{\phi\phi} )^{-1}$.
We limit our analysis to modes $l\!\ge\!150$ of the CIB maps.
We model the power spectrum of the CIB following Ref.~\cite{Addison:2011se}, with $C_l^{{\rm CIB}\mbox{-}{\rm CIB}} = 3500 (l/3000)^{-1.25} {\rm Jy^2 / sr}$.
We model the cross-spectrum $C_l^{{\rm CIB}\mbox{-}\phi}$ between the CIB fluctuations and the lensing potential using the SSED model of Ref.~\cite{hall10}, 
which places the peak of the CIB emissivity at redshift $z_c = 2$ with a broad redshift kernel of width $\sigma_z = 2$.
We choose a linear bias parameter for this model to agree with the results of
Refs.~\cite{Holder:2013hqu,Ade:2013aro}.
More realistic multi-frequency CIB models are available (for example, Ref.~\cite{Bethermin:2013nza});
however, we only require a reasonable template.
The detection significance is independent of errors in the amplitude of the assumed ${\rm CIB}\mbox{-}\phi$ correlation.

\textbf{Results:}
In Fig.~\ref{fig:map_ebp_tri}, we plot Wiener-filtered estimates $\hat{E}^{150}$ and $\hat{\phi}^{\rm CIB}$ using the 
CMB measured by SPTpol at 150\,GHz and the CIB fluctuations traced by 
\textit{Herschel}.
In addition, we plot our estimate of the lensing $B$ modes obtained by applying Eq.~\eqref{eqn:blen} to these measurements.
\begin{figure}[!]
\centerline{\includegraphics[width=\columnwidth]{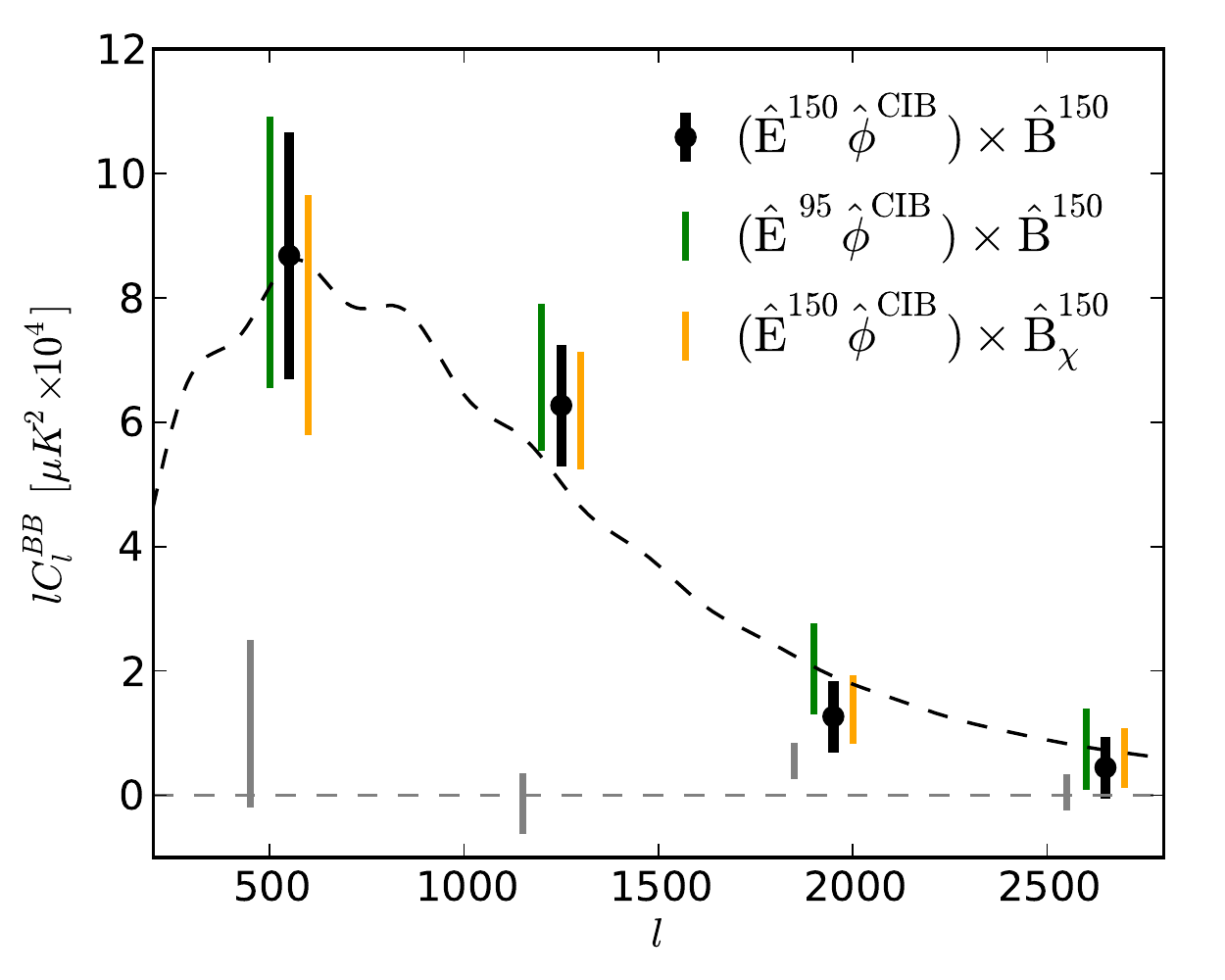}}
\vspace{-0.2in}
\caption{
Black, center bars: cross correlation of the lensing $B$ modes measured by SPTpol 
at $150$\,GHz with lensing $B$ modes inferred from 
CIB fluctuations measured by \textit{Herschel} and
$E$ modes measured by SPTpol at $150\,$GHz; as shown in Fig.~\ref{fig:map_ebp_tri}.
Green, left-offset bars: 
same as black, but using $E$ modes measured at $95$\,GHz,
testing both foreground contamination and instrumental systematics.
Orange, right-offset bars: 
same as black, but with $B$ modes obtained using the $\chi_B$ procedure described in the text rather than our fiducial Wiener filter.
Gray bars: curl-mode null test as described in the text.
Dashed black curve: lensing $B$-mode power spectrum in the fiducial cosmological model.
}
\label{fig:plot_cl_ep_b_pub}
\end{figure}
In Fig.~\ref{fig:plot_cl_ep_b_pub} we show the cross-spectrum between this lensing $B$-mode estimate 
and the $B$ modes measured directly by SPTpol.
The data points are a good fit to the expected cross-correlation, with a $\chi^2/{\rm dof}$ of 
$3.5/4$ and a corresponding probability-to-exceed (PTE) of 
$48\%$.
We determine the uncertainty and normalization of the cross-spectrum estimate using an ensemble of
simulated, lensed CMB+noise maps and simulated \textit{Herschel} maps.
We obtain comparable uncertainties if we replace any of the three fields involved in this procedure with observed data rather than a simulation, 
and the normalization we determine for each bin is within $15\%$
of an analytical prediction based on approximating the Wiener filtering procedure as diagonal in Fourier space.

\begin{figure}
\centerline{\includegraphics[width=\columnwidth]{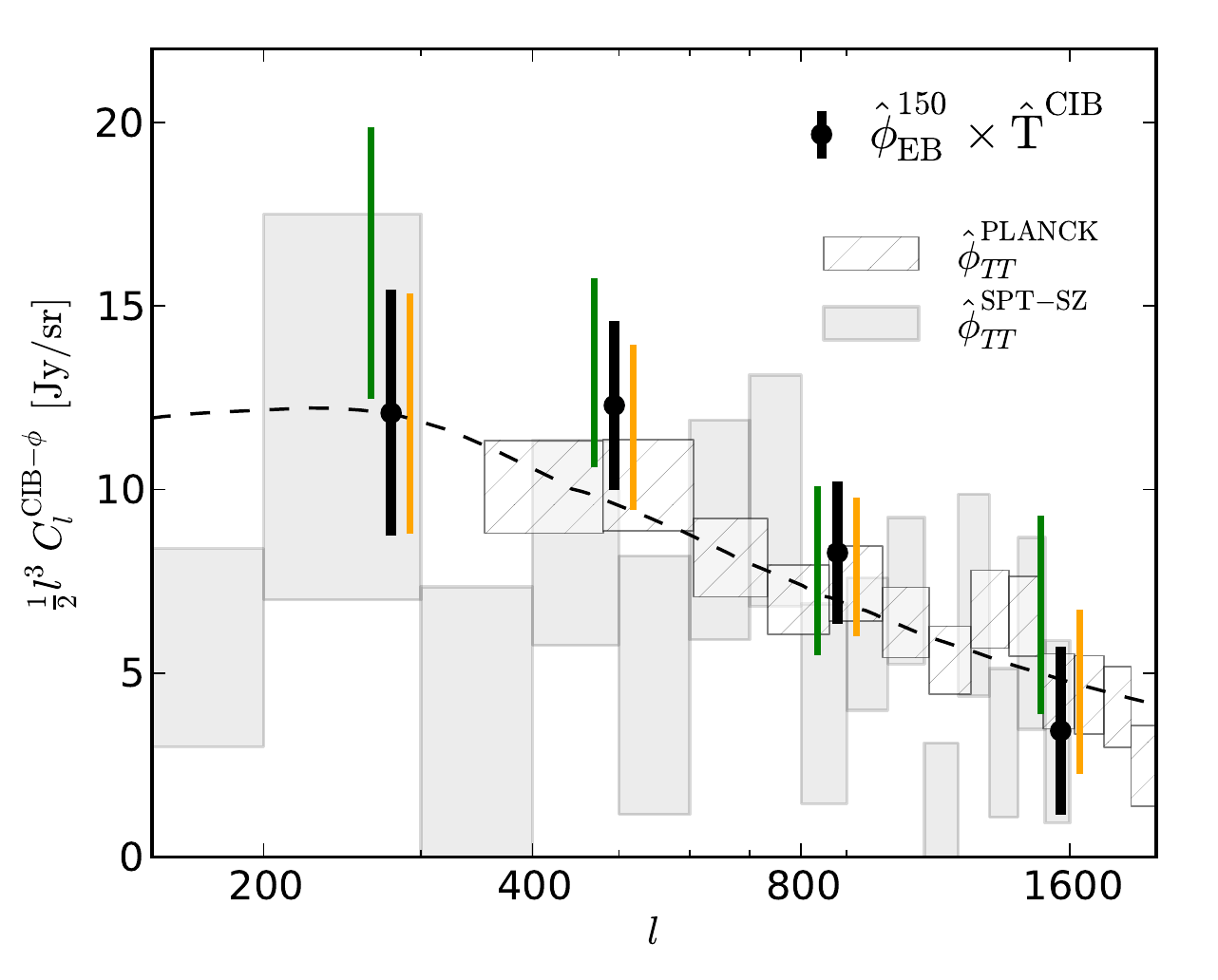}}
\vspace{-0.2in}
\caption{``Lensing view'' of the $EB\phi$ correlation plotted in Fig.~\ref{fig:plot_cl_ep_b_pub}, in which we cross-correlate an $EB$ lens reconstruction from SPTpol data with CIB intensity fluctuations measured by \textit{Herschel}.
Left green, center black, and right orange bars are as described in Fig.~\ref{fig:plot_cl_ep_b_pub}.
Previous analyses using temperature-based lens reconstruction from \textit{Planck} \cite{Ade:2013aro} and SPT-SZ \cite{Holder:2013hqu} are shown with boxes.
The results of Ref.~\cite{Ade:2013aro} are at a nominal wavelength of $550\,\micron$, 
which we scale to $500\,\micron$ with a factor of $1.22$ \cite{Gispert:2000np}.
The dashed black curve gives our fiducial model for $C_l^{{\rm CIB}\mbox{-}\phi}$ as described in the text.}
\label{fig:plot_cl_eb_p_pub}
\end{figure}
In addition to the cross-correlation 
$E\phi\!\times\!B$, it is also interesting to take a ``lensing perspective'' and rearrange the fields to measure the correlation 
$EB\!\times\!\phi$.
In this approach, we perform a quadratic ``EB'' lens reconstruction \cite{Hu:2001kj} 
to estimate the lensing potential $\hat{\phi}_{EB}$, which we then cross-correlate with CIB fluctuations.
The observed cross-spectrum can be compared to previous temperature-based lens reconstruction results \cite{Holder:2013hqu,Ade:2013aro}.
This cross-correlation is plotted in Fig.~\ref{fig:plot_cl_eb_p_pub}.
Again, the shape of the cross-correlation which we observe is in good agreement with the fiducial model, with a $\chi^2/{\rm dof}$ of 
$2.2/4$ and a PTE of $70\%$.

Both the 
$E\phi\!\times\!B$ and 
$EB\!\times\!\phi$ cross-spectra discussed above are probing the three-point correlation function (or bispectrum) between 
$E$, $B$, and $\phi$ that is induced by lensing.
We assess the overall significance of the measurement by constructing a 
minimum-variance estimator for the amplitude $\hat{A}$ of this bispectrum,
normalized to have a value of unity for the fiducial cosmology+CIB model 
(analogous to the analyses of Refs.~\cite{Smith:2007rg,Hirata:2008cb} for the $TT\phi$ bispectrum).
This estimator can be written as a weighted sum over either of the two cross-spectra already discussed.
Use of $\hat{A}$ removes an arbitrary choice between the lensing or $B$-mode perspectives, as both are simply collapsed faces of the $EB\phi$ bispectrum. 
Relative to our fiducial model, we measure a bispectrum amplitude
$\hat{A} = 1.092 \pm 0.141$, 
non-zero at approximately
$7.7\sigma$.

We have tested that this result is insensitive to analysis choices.
Replacement of the $B$ modes obtained using the baseline Wiener filter
with those determined using the $\chi_B$ estimator causes a shift of $0.2\sigma$.
Our standard $B$-mode estimate incorporates a mask to exclude bright point sources, while the $\chi_B$ estimate does not. 
The good agreement between them indicates the insensitivity of polarization lensing measurements to point-source contamination.
If we change the scan direction cut from $l_x\!<\!400$ to $200$ or $600$, the measured
amplitude shifts are less than $1.2\sigma$, consistent with the root-mean-squared 
(rms) shifts seen in simulations.
If we repeat the analysis without correcting for 
\mbox{I $\rightarrow$ Q,\,U} leakage, the measured amplitude shifts by less than $0.1\sigma$.
A similar shift is found if we rotate the map polarization vectors by one degree to mimic 
the $E\rightarrow B$ leakage which would be induced by 
an error
in the average PSB angle.

We have produced estimates of $\hat{B}^{\rm lens}$ using alternative estimators of $E$. 
When we replace the $E$ modes measured at $150$\,GHz with those measured at $95$\,GHz, 
we measure an amplitude \mbox{$\hat{A} = 1.225 \pm 0.164$}, 
indicating a lack of significant foreground contamination or spurious correlations between $150$\,GHz $E$ and $B$ modes.
We have also estimated the bispectrum amplitude using the CMB temperature as a tracer of $E$ (exploiting the $C_l^{TE}$ correlation, 
as in Refs.~\cite{Jaffe:2003ik,Dore:2003wp}).
We obtain an amplitude of 
$\hat{A} = 1.374 \pm 0.427$, consistent with unity.
The shifts between these values of $\hat{A}$ and our fiducial one are consistent with the rms scatter seen in simulations.

Our amplitude estimate also passes several null tests.
Using $E$ modes obtained from the difference map of left-going and right-going scans,  we measure
\mbox{$\hat{A} =  0.049 \pm 0.037$.}
Replacing instead the $B$ modes with a difference map, we obtain
\mbox{$\hat{A} = -0.028 \pm 0.119$.}
We also estimate the amplitude of the $B$-mode power sourced by curl-type lensing modes.
These generate deflections as the curl of a scalar potential rather than a gradient \cite{Cooray:2005hm}.
Curl-type lensing modes are expected to be negligibly small in the fiducial cosmological model \cite{Hirata:2003ka}, 
but could be sourced by instrumental effects.
If we take the CIB as a tracer of curl-type lensing modes and measure the lensing $B$-mode power in cross-correlation
analogously, we obtain the gray errorbars in Fig.~\ref{fig:plot_cl_ep_b_pub}, which are consistent with the expected value of zero,
having a $\chi^2/{\rm dof}$ of $2.9/4$ (PTE = $57\%$).

As a final consistency test, we also replace the CIB-based estimate of $\phi$
with quadratic lens reconstructions based on combinations 
$TT$, $TE$, $EE$, and $EB$ of the CMB fluctuations \cite{Guzik:2000ju,Benabed:2000jt,Hu:2001kj,Hirata:2003ka}.
These lensing estimators probe the CMB lensing potential directly, 
and are not limited by the imperfect overlap of the CIB redshift kernel with that of lensing.
The temperature-only lensing estimator has already been used to make precise measurements of the
lensing potential power spectrum \cite{Das:2011ak,vanEngelen:2012va,Das:2013zf,Ade:2013tyw}.
We now have the ability to extend this analysis to polarization.
We have not yet performed a thorough characterization of these estimators or their sensitivity to
analysis choices and possible systematic effects in the SPTpol data; 
however, comparing the measured amplitudes with our fiducial result using statistical error bars is still a useful consistency test.
We obtain amplitude estimates in reasonable agreement with the fiducial cosmological model, with 
\mbox{$\hat{A}^{TT}\!=\!0.675 \pm 0.194$}, 
\mbox{$\hat{A}^{TE}\!=\!0.921 \pm 0.294$},
\mbox{$\hat{A}^{EE}\!=\!0.860 \pm 0.267$},
and
\mbox{$\hat{A}^{EB} = 0.960 \pm 0.386$}.
These estimators probe the four-point correlations (or trispectra) of the lensed CMB, which can be non-zero even for Gaussian fluctuations.
We have subtracted a ``bias-hardened'' estimate of the Gaussian contribution \cite{Namikawa:2012pe} for each estimator to correct for this.

\textbf{Conclusions:}
$B$-mode polarization is a promising avenue for measuring
both the CMB lensing potential $\phi$ and primordial gravitational waves.
In this Letter, we have presented the first detection of $B$ modes produced by gravitational lensing.
Beyond this detection, the lensing $B$-mode map which we have synthesized can be subtracted from the
observed $B$ modes in a process of ``delensing'', reducing the effective noise level for the measurement of primordial $B$ modes \cite{Knox:2002pe,Kesden:2002ku,Seljak:2003pn,Smith:2008an}.
The work presented here is a first step in the
eventual exploitation of CMB $B$-mode polarization as a probe of
both structure formation and the inflationary epoch.

\ifwordcount
\else
\acknowledgements
\textbf{Acknowledgements:}
The SPT is supported by the National Science Foundation through grant ANT-0638937, with partial support provided by NSF grant PHY-1125897.
Support for the development and construction of SPTpol were provided by the Gordon and Betty Moore Foundation through Grant GBMF 947 to the University of Chicago, a gift from the Kavli Foundation, and NSF award 0959620.
\textit{Herschel} is an ESA space observatory with science instruments provided by European-led Principal Investigator consortia and with important participation from NASA.
This research used resources of the National Energy Research Scientific Computing Center, which is supported by the Office of Science of the U.S. Department of Energy under Contract No. DE-AC02-05CH11231.
It also used resources of the CLUMEQ supercomputing consortium, part of the Compute Canada network.
Research at Argonne National Laboratory and use of the Center for Nanoscale Materials are supported by the Office of Science of the U.S. Department of Energy under Contract DE-AC02-06CH11357.
The McGill group acknowledges funding from the National Sciences and Engineering Research Council of Canada, Canada Research Chairs program, and the Canadian Institute for Advanced Research.
The C.U.\ Boulder group acknowledges support from NSF AST-0956135. We thank P.\ Hargrave at Cardiff University for anti-reflection coating the SPTpol lens, A.\ Datesman for his work on TES detectors at Argonne,  R.\ Divan for microfabrication support at Argonne, and the members of the Truce collaboration for their efforts in the design of the 150 GHz polarization detectors, in particular D.\ Becker, J.\ Britton, M.D.\ Niemack, and K.W.\ Yoon at NIST. We thank M.\ Lueker, T.\ Plagge, Z.\ Staniszewski, E.\ Shirokoff, H.\ Spieler and R.\ Williamson for their considerable contributions to the SPT program.
DH was supported by the Lorne Trottier Chair in Astrophysics and Cosmology at McGill as well as a CITA National Fellowship.
RK acknowledges support from NASA Hubble Fellowship Grant HF-51275.
\bibliography{ebp_paper,spt}
\fi

\end{document}